\documentstyle[11pt]{article}

\begin{document}

\def\be{\begin{equation}}
\def\ee{\end{equation}}
                         \def\eg{ {\em e.g.}}
                         \def\etc{ {\em etc.}}
                         \def\etal{ {\em et al.}}
                         \def\ie{ {\em i.e.}}
                         \def\viz{ {\em viz.}}
\def\lsim{\:\raisebox{-0.5ex}{$\stackrel{\textstyle<}{\sim}$}\:}
\def\gsim{\:\raisebox{-0.5ex}{$\stackrel{\textstyle>}{\sim}$}\:}
                         \def\go{\rightarrow}
                         \def\goes{\longrightarrow}
\def\rpv{R \!\!\!\!/}
\def\mET{E_T \!\!\!\!\!\!\!/~~}

\def\N0{\tilde \chi^0}
\def\Cp{\tilde \chi^+}
\def\Cm{\tilde \chi^-}
\def\Cpm{\tilde \chi^\pm}

\def\lpp{\lambda''}
\def\lp{\lambda'}
\def\l{\lambda}
\def\tb{\tan\beta}

\def\sq{\tilde q}
                         \def\su{\tilde u}
                         \def\sd{\tilde d}
                         \def\sc{\tilde c}
                         \def\ss{\tilde s}
                         \def\st{\tilde t}
                         \def\sb{\tilde b}

\def\sl{\tilde \ell}
                         \def\sel{\tilde e}
                         \def\snu{\tilde \nu}
                         \def\smu{\tilde \mu}
                         \def\stau{\tilde \tau}

\def\emem{e^- e^-}
\def\epem{e^+ e^-}
\def\egam{e \gamma}
\def\gamgam{\gamma \gamma}

\setcounter{page}{0}
\thispagestyle{empty}

\begin{flushright}
                   CERN-TH/97-341 \\
                   hep-ph/9711473
\end{flushright}

\begin{center}

{\Large\bf    
$R$-parity Violation in Neutralino Decays at an $e\gamma$ Collider 
} \\

\vskip 30pt

{\bf 
      Dilip Kumar Ghosh 
     \footnote{Current address: Centre for Theoretical Studies,
                                Indian Institute of Science,
                                Bangalore 560 012, India. 
                                E-mail: dghosh@cts.iisc.ernet.in}
} \\

{\footnotesize\rm 
     Department of Physics, University of Mumbai,\\
     Vidyanagari, Santa Cruz (East), Mumbai 400 098, India.
} \\

\bigskip\bigskip

{\bf 
      Sreerup Raychaudhuri \footnote{E-mail: sreerup@mail.cern.ch}
} \\

{\footnotesize\rm 
     Theory Division, CERN, CH 1211 Geneva 23, Switzerland.
} \\

\vskip 30pt

{\bf 
    ABSTRACT
}

\end{center}

\begin{quotation}

At an $\egam$ collider, a selectron $\sel_{L,R}$ may be produced in 
association with a (lightest) neutralino $\N0_1$. Decay of the selectron 
may be expected to yield a final state with an electron and another 
$\N0_1$. If $R$-parity is violated, these two neutralinos will decay,
giving rise to distinctive signatures, which are identified and studied.

\end{quotation}

\vskip 50pt

\begin{flushleft}
                   CERN-TH/97-341 \\
                   November 1997
\end{flushleft}


\newpage

For a variety of theoretical and phenomenological reasons, of which the
unification of strong and electroweak couplings at a common scale is
perhaps the most exciting \cite{SUSY_Unif}, supersymmetry is currently the 
most popular option for going beyond the Standard Model (SM) of strong 
and electroweak interactions. Of course, even the {\em minimal} extension 
of the SM which incorporates supersymmetry (MSSM) predicts a large number 
of new particles and interactions which have not (yet) been seen. One of 
the major goals for particle physics is, thus, to study different
strategies for the detection of these new particles. Many such analyses
have already appeared in the literature \cite{SUSY_Revs}. It is 
apparent, however, that the current generation of high energy accelerators
has already exhausted a certain part of its potential for new particle 
discovery and it is entirely possible that we will end up with a set of 
useful, but uninspiring, bounds. One would therefore have to achieve higher 
energies or alternative techniques if indeed the new physics --- whether 
supersymmetry or something else --- is to be found.

It is partly with this kind of future scenario in mind that several plans
for new accelerators and 
accelerator techniques have been suggested, many of which are
now under serious consideration. One of the most interesting of these is
the suggestion of a high energy $e\gamma$ collider, which would be the
first machine of its kind. 
The basic plan is to have a high energy
$\epem$ (or $\emem$) collider (such as the 500 GeV NLC, for instance) 
and to direct
a highly coherent laser beam at the positron (electron) beam at small
angles; the
back-scattered laser beam, which picks up most of the energy of the
positron (electron) beam, could then be allowed to collide with the
other electron (positron) beam, leading to $\egam$ interactions at
high energies. This involves no new principle beyond inverse
Compton scattering and detailed studies of the design and properties of
such a machine have already appeared in the literature \cite{EGam_Tech}. 
There have
also been quite a few explorations of its physics possibilities,
especially as a probe of low-energy supersymmetry \cite{EGam_Phys}.

One of the most interesting of these physics possibilities 
is the production of single
selectrons through the process $\egam \go \sel_{L,R} ~+~ \N0_i ~(i =
1,2,3,4)$. This can occur through an $s$-channel electron exchange or
a $t$-channel selectron exchange. If the selectron 
decays to an electron (positron) and another neutralino, then we are
left with an $e^- \N0_i \N0_j$ final state. 
We concentrate on the $e^- \N0_1 \N0_1$ state, which is kinematically 
favoured. In
the canonical form of the MSSM, $R$-parity is conserved and the
$\N0_1$ is the best candidate for the lightest superymmetric particle
(LSP), which does not decay and which escapes the detectors. The
signal for this process is, then, a single hard electron and 
substantial 
missing energy and momentum. This process, together with its obvious
SM background from $\egam \go e \nu \bar \nu$, has been discussed by 
several authors as a possible signal for supersymmetry 
\cite{EGam_Phys,Choudhury,Barger}.

Though it was pointed out long ago \cite{SUSY_RpV} that $R$-parity 
conservation in
the MSSM is not really demanded by any theoretical or experimental
considerations, this rather less-than-elegant alternative was not taken very
seriously during the first decade of studies in supersymmetry.
Recently, in the wake of the announcement of an excess observed 
in high-$Q^2$
$ep$-scattering events at HERA \cite{HERA}, there has been a resurgence of
interest in the possibility of $R$-parity violation, since this seems
to provide a natural explanation of the excess in terms of a squark
resonance \cite{HERA_RpV}. While the status of the HERA excess is 
still uncertain, what {\em is} certain is that $R$-parity violation is 
increasingly being recognised as an alternative 
scenario to the canonical form of the MSSM where $R$-parity is conserved.
It is, thus, important to study the experimental consequences of $R$-parity
violation, especially in the context of present and future high energy
colliders, including the $\egam$ collider discussed above.

In this letter, we consider, therefore, the possibility that
$R$-parity is not conserved and that the lightest neutralino (LSP)
can, as a
consequence, decay into three-fermion final states, whose flavour content
depends on the nature of the $R$-parity-violating interactions. The
selectron-neutralino 
production process will then lead to rather spectacular
multi-fermion final states, which should have very little SM
backgrounds. The observation of such states could be a clear
signal for superymmetry. We restrict our discussion
to the so-called {\em weak} limit, in which
$R$-parity-violating couplings are small compared to gauge couplings
and thus most production and decay mechanisms of supersymmetric
particles occur exactly as in the canonical
supersymmetric models, the only new feature being the decay of the 
neutralino LSP. If we were to allow somewhat larger $R$-parity-violating
couplings --- which are not experimentally disallowed \cite{Dreiner} ---
the scenario can change quite significantly. Even in the
weak $R$-parity violation scenario,
production of a selectron-neutralino pair is not the only
process at an $\egam$ collider where $R$-parity-violating effects 
could appear. One can also have such effects in the decays following
production of a sneutrino-chargino pair \cite{Barger}. 
Moreover, in $R$-parity-violating models, it is also conceivable that 
the {\em selectron}, rather than the lightest neutralino, is the LSP, and
can decay directly into two leptons (quarks) through the 
$R$-parity-violating $\l (\lp)$ coupling. For larger $\l$ 
couplings, there is also the possibility
of producing lepton-flavour-violating $\ell_i \snu_j$ final states 
directly. None of these signals are considered in this letter. 
A more 
comprehensive study of these processes is certainly called for and will
be taken up in a forthcoming publication \cite{selves}.

The basic processes leading to the production of a
selectron-neutralino final state have already been discussed in the
literature. Cross-sections have been listed in Ref.
\cite{Barger} for polarised beams. The photon flux and polarisation 
have been worked out by Ginzburg
and his collaborators \cite{EGam_Tech} and are quoted in Refs.
\cite{EGam_Phys,Choudhury}. In the interests of brevity, we do not
reproduce all these formulae here and refer the reader to the
existing literature. We simply mention the assumptions we have made
in our numerical analysis. Briefly, these are the following:
\begin{enumerate}
\item
We assume that the laser back-scattering parameter $x = 2(1 +
\sqrt{2}) \simeq 4.828$, its maximum value \cite{EGam_Tech}; 
\item
It is likely that the relatively low-energy photons in the beam 
would be lost \cite{EGam_Tech}
and thus the momentum fraction $y = E_\gamma/E_e$ of the photon beam lies 
between
0.5 and the maximum value $x/(1+x) \simeq 0.828$ (this is identical
with the choice of Ref.\cite{Choudhury}); 
\item
The energy of the initial electron and positron beams are taken to be
250 GeV each (which is expected to be available at the NLC); 
\item
The polarisation of the initial laser beam ($P_l$), the 
initial positron beam
($\l_p$) which scatters the laser and the initial unscattered
electron beam ($\l_e$) are taken to be $|P_l| = 1.0, |\l_p| = 0.4, \l_e =
\pm 0.45$; the signs of $P_l, \l_p$  are consistent with the choices 
of Ref. \cite{Barger}.
\item
With these more-or-less optimal choices of parameters, the luminosity 
expected for the $\egam$ collider would probably be of the same order
as that of the parent $\epem$ machine \cite{Choudhury,Barger}. 
We assume a representative value of 10 fb$^{-1}$.
\end{enumerate}

\begin{figure}[h]
\begin{center}
\vspace {5.2in}
      \relax\noindent\hskip -3.5in\relax{\includegraphics{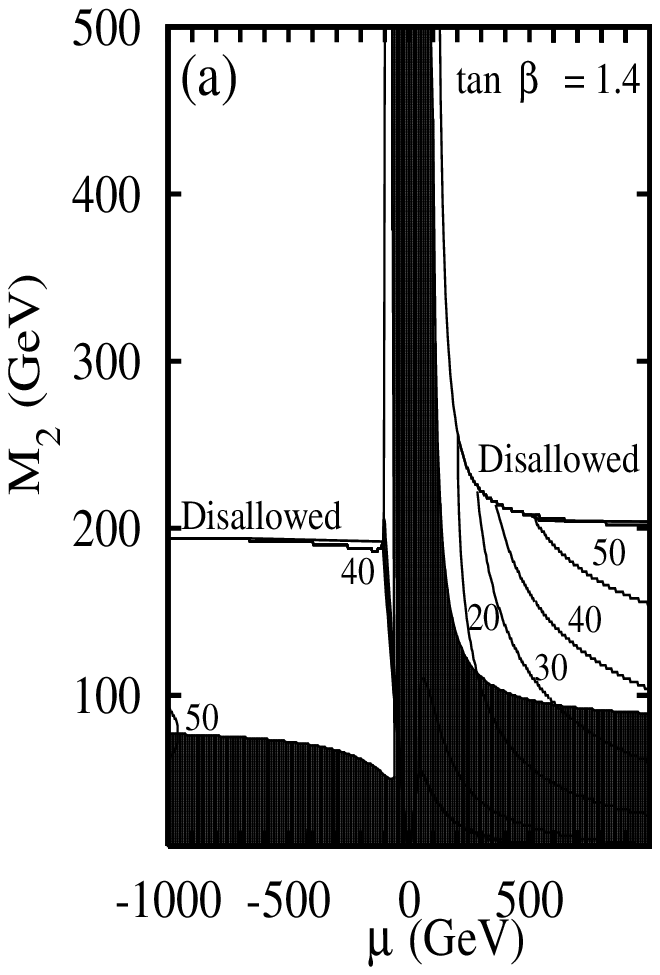}}
      \hskip 2.0in\relax{\includegraphics{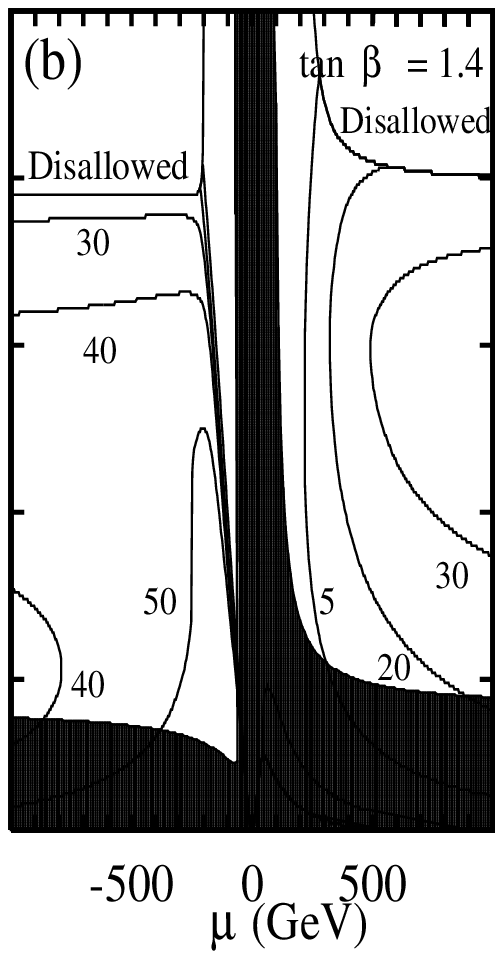}}
      \hskip 2.0in\relax{\includegraphics{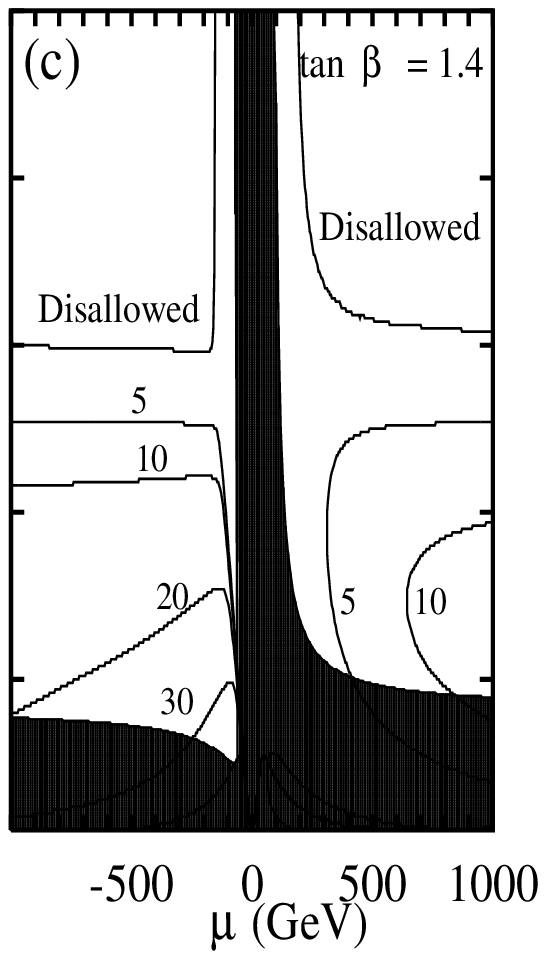}}
\end{center}
\vspace*{-120pt}
\caption{\footnotesize\em
           Contours of cross-section (marked, in fb) 
           for a left selectron of mass 
           (a) 100 GeV, (b) 200 GeV and (c) 300 GeV. We choose 
           the polarisation of the electron beam to be 
           $\lambda_e = -0.45$. 
           Larger cross-sections 
           lie away from the origin for $\mu > 0$ 
           and towards the origin for $\mu < 0$.
        }
\end{figure}

With the above assumptions we have incorporated the relevant formulae 
into a simple parton-level Monte Carlo 
event generator. In Fig. 1 we present contour plots of the total 
cross-section for $\sel_L \N0_1$ production in the 
$M_2-\mu$ plane for $\tb = 1.4$ and three different values of 
{\em left}-selectron mass $m_{\sel_L} = (a)~100,~(b)~200,~{\rm and}~(c)~300$ 
GeV respectively. We assume gaugino mass unification (at least in the 
electroweak sector) throughout this discussion. 
The cross-section (in fb) is marked next to the
relevant contours. The region marked `Disallowed'
represents the region where either ($a$) the production of a 
selectron-neutralino
pair is kinematically disallowed, or 
($b$) the selectron becomes the LSP and 
cannot decay to a neutralino \footnote{Of course, one
could still look for $R$-parity violation in decays of the selectrons
(for the neutralino produced in association with the selectron
may decay into an additional selectron), but this interesting
scenario is beyond the scope of this letter.}. These contours are the
same as those we would get in a model with conserved $R$-parity. 
The shaded region is ruled out 
by direct searches at LEP-2 for $R$-parity violation from an 
$LQ\bar D$-type operator, 
and agrees closely with the kinematic limit for
chargino production. 
It is worth mentioning that the value of $\tan \beta$ chosen
for this figure is at the edge of the allowed region (from the LEP Higgs
search constraints) and is 
principally chosen because ($a$) this minimises the ruled-out region 
in the $M_2 - \mu$ plane, and
($b$) this is the common choice made by the LEP collaborations, so that
the ruled-out region may be read off from their plots as well 
\footnote{For $LL\bar E$-operators, care should be taken in reading off the
results presented by the LEP collaborations \cite{LEP}, since they assume 
somewhat higher slepton masses than are considered in  Figs. 1 and 2.}.
We also note that our predictions of the cross-section agree well with those 
of Refs. \cite{EGam_Phys,Choudhury,Barger}.

It is immediately obvious that, for a projected luminosity of 10 fb$^{-1}$, 
one can obtain a few hundreds
 of left selectron events over a respectable range of parameter space, with 
negative values of $\mu$ being distinctly preferred. 
For a selectron mass of 100 
GeV, the parameter space of interest is restricted by the requirement 
that the neutralino be the LSP, which forces it to be
 lighter than 100 GeV in this case.
For $m_{\sel_L} = 300$ GeV, the parameter space is again restricted by the
machine energy limitations, while
 the cross-section itself is limited by phase-space
considerations. A left selectron mass in the range of 200 GeV, however, 
seems to be optimum for the $\egam$ collider, as Fig. 1($b$) makes clear.

In Fig. 2, we present the analogue of Fig. 1 (with the same notations
and conventions), for the {\em right} selectron. However, in determining
these cross-sections, the sign of the electron
beam polarisation 
has been reversed, since the earlier polarisation was designed specifically
to produce the left selectron to the exclusion of the right. 
It is obvious that, other things being equal, the cross-sections
are significantly larger than those of Fig. 1, primarily because 
the larger hypercharge of the right selectron couples 
it more strongly to the bino component of the neutralino (this is 
dominant over most of the parameter
space for low values of $\tb$). Thus a few thousand right
selectrons could easily be produced in 10 fb$^{-1}$ of data, making
detection a relatively simple matter.

\begin{figure}[h]
\begin{center}
\vspace {5.2in}
      \relax\noindent\hskip -3.5in\relax{\includegraphics{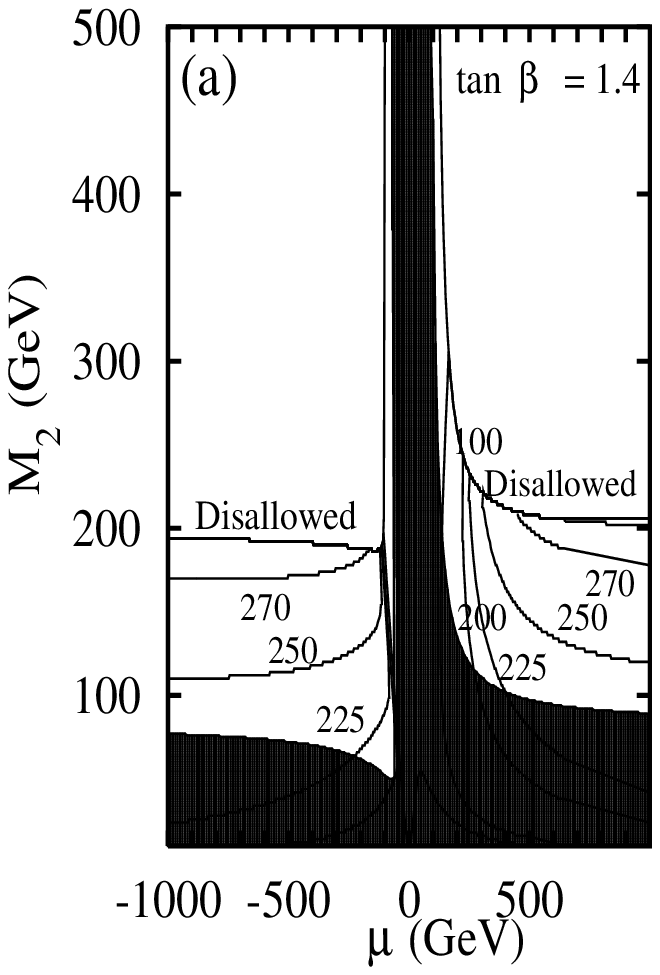}}
      \hskip 2.0in\relax{\includegraphics{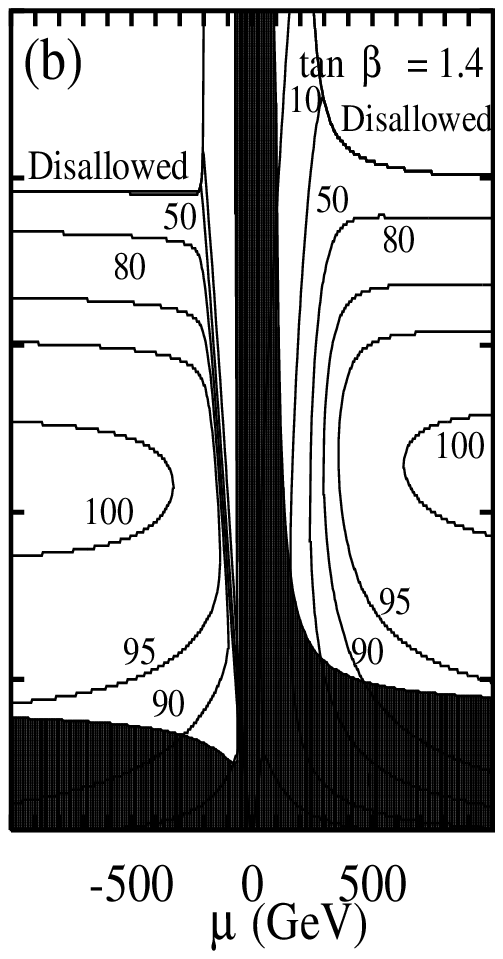}}
      \hskip 2.0in\relax{\includegraphics{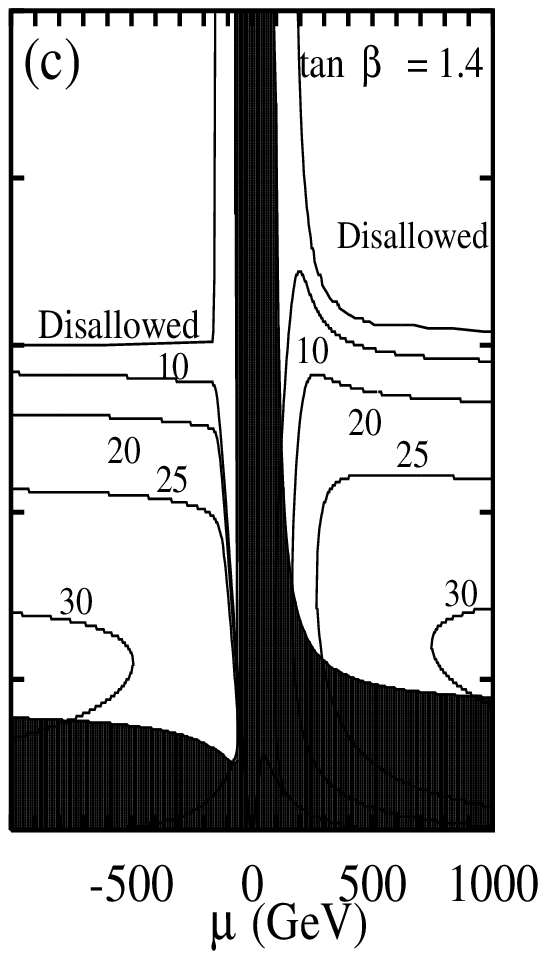}}
\end{center}
\vspace*{-120pt}
\caption{\footnotesize\em 
           Contours of cross-section (marked, in fb) 
           for a right selectron of mass
           (a) 100 GeV, (b) 200 GeV and (c) 300 GeV. We choose
           the polarisation of the electron beam $\lambda_e = +0.45$.
           Larger cross-sections (marked, in fb)
           lie away from the origin for both
           $\mu < 0$ and $\mu > 0$.
        }
\end{figure}

The selectron can now decay in any one of the following 
{\em three} channnels: ($a$) into an electron and a neutralino; ($b$) into
a neutrino and a chargino (left selectron only); 
($c$) into a neutrino and a charged lepton or
into a pair of quarks (depending on the $R$-parity-violating coupling). 
For small values of $\l$ or $\lp$ (such as are assumed 
here) the last mode does not 
contribute much (except at the very edge of the parameter space where the 
electron-neutralino mode gets kinematically suppressed) and will not be 
considered any further. 
The neutrino-chargino mode, is, however, important for the left selectron, 
and though we do not consider the corresponding final states in this
letter, we must take into account the reduction in the branching
ratio to electron-neutralino final states. In Fig. 3, we plot contours of
the branching ratio of a 
left selectron of mass 200 GeV 
into the electron-neutralino channel in the
$M_2-\mu$ plane for $\tb = 1.4$, 
assuming a vanishing contribution from the $R$-parity-violating
channel(s). The parameters for Fig. 3 are chosen so that it is
possible to convolute the branching ratio with production cross-sections
read-off from Fig. 1($b$). For the right selectron, under similar 
assumptions, we get a 100 \% branching ratio to
the electron-neutralino mode. Apart from the branching
ratio, the decay of the selectron into an electron and a neutralino 
should lead to a hard central electron which we identify by putting the 
following cuts: ($a$) pseudorapidity $|\eta_e| < 3$ and ($b$) 
transverse energy $E_T^e > 10$ GeV. These cuts, as may be expected,
have very little effect 
on the cross-section, the reduction being by 1\% or less.

\bigskip

\begin{figure}[h]
\begin{center}
\vspace {4.2in}
      \relax\noindent\hskip -4.0in\relax{\includegraphics{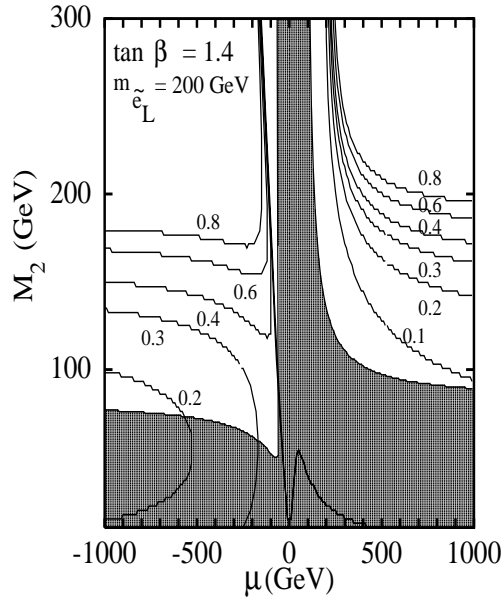}}
\end{center}
\vspace*{-120pt}
\caption{\footnotesize\em Contours of branching ratio for a left selectron
                          to electron and neutralino.
        }
\end{figure}

Till now, most of the results and discussions have 
not deviated from the $R$-parity-conserving case. 
We now turn to the decay modes of the 
neutralinos (LSP's) where much depends on
the assumptions made about the $R$-parity-violating sector. The most
general such interaction arises from the superpotential
\be
{\cal W}_{\rpv} = 
\l_{ijk}\epsilon_{\alpha\beta}
{\hat L_i^\alpha}{\hat L_j^\beta} {\hat {\bar E_k}}
+\lp_{ijk}\epsilon_{\alpha\beta}
{\hat L_i^\alpha}{\hat Q_j^\beta}{\hat {\bar D_k}}
+ \mu_{ij}\epsilon_{\alpha\beta}
{\hat L_i^\alpha}{\hat H_j^\beta}
+\lpp_{ijk}
{\hat {\bar U_i}}{\hat {\bar D_j}}{\hat {\bar D_k}}
\ee
where ${\hat L},{\hat Q},{\hat H}$ 
are $SU(2)$ doublets containing lepton, quark and Higgs
superfields respectively and all colour indices have been dropped.
The $SU(2)$ structure demands that $\l$ be antisymmetric in the first
two indices and $SU(3)_c$ invariance demands that $\lpp$ be
antisymmetric in the last two. The bilinear couplings can be absorbed in the
corresponding trilinear ones by a redefinition of leptonic superfields
and will play no further role in the subsequent discussion. 
The first three terms in the
superpotential violate lepton number (but not baryon number) and the
last one violates baryon number (but not lepton number).

One can now expand the $F$-term of the above superpotential to get
the interaction vertices.
We note that most (though not quite all) 
products of lepton-number-violating couplings of the $LQ\bar D$-type 
with baryon-number-violating 
couplings are severely constrained by the observed stability of the proton
\cite{Dreiner,proton_decay}. 
A simple expedient to satisfy these constraints is to allow
nonconservation of either lepton number, or of baryon number, but not of both. 
In most of 
the following discussion, we shall assume that only {\em lepton} number is 
violated. This option is, in fact, reasonably 
well-motivated from a theoretical point of view \cite{Ross}. In fact,
even for the lepton-number-violating couplings, we shall study
signals assuming that {\em one} coupling at a time dominates. This is
a rather naive assumption, but it is convenient and, at this 
exploratory stage of our study, seems a reasonable one to make.
After all, one can argue that 
the situation is similar for the SM Yukawa couplings where 
the coupling of the top quark is overwhelmingly larger than the others.

We first concentrate on the $LL\bar E$ operators. In the presence of a
$\l_{ijk}$ coupling, the neutralino LSP undergoes the three-body decays
\begin{displaymath}
\N0_1 \goes \nu_i \ell_j^- \ell_k^+ + \nu_j \ell_i^- \ell_k^+
\end{displaymath}
through a virtual $\snu_{i(j)}, \sl_{Lj(i)}, \sl_{Rk}$ exchange
\footnote{Since we assume the $\N0_1$ to be the LSP, the
possibility of slepton resonances in the neutralino decay may be 
discounted.}.
Detailed formulae for these decays have appeared before 
\cite{Morawitz}
in the literature, and hence, are not presented here.
It is adequate for
our purposes to observe that the neutralino decay will
yield a pair of charged leptons, not necessarily of the same flavour,
and missing energy from the escaping neutrino. The Majorana nature of
the $\N0_1$ also ensures that for every $\ell_j^- \ell_k^+$ pair,
observed, there will be a $\ell_j^+ \ell_k^-$ pair as well, with missing 
energy as before.
The possible final states arising from a selectron and a
neutralino together with the $\l$ coupling responsible are listed
in Table 1. We have chosen, for illustration, 
a particular point in the parameter
space ($M_2 = 180$ GeV, $\mu = -500$ GeV, $\tb = 1.4$, $m_{\sel} 
= 200$ GeV) where the cross-sections are relatively 
large (46.14 fb and 114.83 fb for the left and right selectrons
respectively) and the branching 
ratio of the left selectron to the neutralino decay mode is
also large (0.89). 
Though the dependence of the kinematic distributions and hence the
effect of the cuts on the sfermion mass spectrum is minimal, we have
chosen, to be specific, all soft supersymmetry-breaking sfermion mass
parameters to be 500 GeV, except the ones which directly enter into the
analysis with different values (such as the selectron masses and hence
the mass of the $\snu_e$ which is related to the $\sel_L$ mass). We also
set all trilinear couplings to zero, which makes one stop rather lighter
than 500 GeV, but still heavier than the neutralino LSP.

In Table 1, 
only the {\em charged} lepton content of the final state is shown.
Such signals involving five charged leptons and substantial missing
(transverse) energy ($\mET$) are rather distinctive and should be easily
identifiable. For our parton-level analysis, 
we demand that all the charged leptons satisfy the criteria 
already imposed on the
electron (arising from selectron decay), \viz, $|\eta_\ell| < 3$
and $E_T^\ell > 10$ GeV. We further demand that
if there are $\tau$ leptons in the final state, 
the other leptons should be isolated from the
narrow jets arising from the $\tau$'s; for this we use a simple-minded
cone algorithm with $\Delta R_{\ell\tau} > 0.2$, where 
$\Delta R = \sqrt{\Delta \eta^2 + \Delta \phi^2}$. It is also necessary
to have a corresponding separation between the tau jets. We find that the
effect of these selection criteria is to diminish the signal by about
35\% (55\% if there are $\tau$'s). 
We assume an efficiency of 80\% for the identification of each
$\tau$ as contrasted to 95\% for the identification of an $e$ or a $\mu$.
These efficiencies are based on LEP-2 estimates and may change slightly
for the higher energies expected at an $\egam$ collider. However, they
are in the right ballpark 
\footnote{In fact, if they err, it is on the conservative side.}
and it is apparent from Table 1 that they
should still lead to some tens of events per channel 
for a luminosity of 10 fb$^{-1}$.
Cross-sections in the last column of Table 1 include a combinatoric factor
of 2 because the two neutralinos could decay either way.

\bigskip

\footnotesize
\begin{center}
\begin{tabular}{|c|c|c|c|c|c|c|}
\hline
$\l$ & {\rm leptons} & {\rm C.S.(fb)} & {\rm leptons} & {\rm C.S.(fb)} &
{\rm leptons} & {\rm C.S.(fb)} \\
\hline
121 & $~e^- e^\pm   e^\mp e^\pm   e^\mp$        & 5.56 (13.92) &
      $~e^- e^\pm \mu^\mp e^\pm \mu^\mp$        & 5.56 (13.92) &
      $~e^- e^\pm   e^\mp e^\pm \mu^\mp$        & 11.11 (27.84) \\
\hline
122 & $~e^- e^\pm \mu^\mp e^\pm \mu^\mp$        &  5.56 (13.92) &
      $~e^- \mu^\pm \mu^\mp \mu^\pm \mu^\mp$    &  5.56 (13.92) &
      $~e^- e^\pm e^\mp  \mu^\pm \mu^\mp$       & 11.11 (27.84) \\
\hline
123 & $~e^-   e^\pm \tau^\mp   e^\pm \tau^\mp$  & 3.86 (9.64) &
      $~e^- \mu^\pm \tau^\mp \mu^\pm \tau^\mp$  & 3.86 (9.64) &
      $~e^-   e^\pm \tau^\mp \mu^\pm \tau^\mp$  & 7.72 (19.29) \\
\hline
131 & $~e^-  e^\pm    e^\mp e^\pm    e^\mp$     & 5.44 (13.56) &
      $~e^-  e^\pm \tau^\mp e^\pm \tau^\mp$     & 3.87 (9.64) &
      $~e^-  e^\pm    e^\mp e^\pm \tau^\mp$     & 9.18 (22.83) \\
\hline
132 & $~e^-   e^\pm  \mu^\mp   e^\pm  \mu^\mp$  & 5.44 (13.57) &
      $~e^- \mu^\pm \tau^\mp \mu^\pm \tau^\mp$  & 3.88 (9.63) &
      $~e^-   e^\pm  \mu^\mp \mu^\pm \tau^\mp$  & 9.19 (22.84) \\
\hline
133 & $~e^-    e^\pm \tau^\mp    e^\pm \tau^\mp$ & 3.58 (8.91) &
      $~e^- \tau^\pm \tau^\mp \tau^\pm \tau^\mp$ & 2.55 (6.35) &
      $~e^-    e^\pm \tau^\mp \tau^\pm \tau^\mp$ & 6.05 (15.01) \\
\hline
231 & $~e^- e^\pm  \mu^\mp e^\pm  \mu^\mp$      & 5.44 (13.57) &
      $~e^- e^\pm \tau^\mp e^\pm \tau^\mp$      & 3.87 (9.64) &
      $~e^- e^\pm  \mu^\mp e^\pm \tau^\mp$      & 9.18 (22.84) \\
\hline
232 & $~e^- \mu^\pm  \mu^\mp \mu^\pm  \mu^\mp$  & 5.44 (13.57) &
      $~e^- \mu^\pm \tau^\mp \mu^\pm \tau^\mp$  & 3.88 (9.63) &
      $~e^- \mu^\pm  \mu^\mp \mu^\pm \tau^\mp$  & 9.18 (22.84) \\
\hline
233 & $~e^-  \mu^\pm \tau^\mp  \mu^\pm \tau^\mp$ & 3.58 (8.92) &
      $~e^- \tau^\pm \tau^\mp \tau^\pm \tau^\mp$ & 2.55 (6.34) &
      $~e^-  \mu^\pm \tau^\mp \tau^\pm \tau^\mp$ & 6.06 (15.01) \\
\hline
\end{tabular}
\end{center}
\noindent
{\em Table 1. Final states for different $\l$ couplings with
representative cross-sections for the left selectron in each channel. 
Numbers in parantheses
show cross-sections for the right selectron. All (four) possible 
sign-combinations of the final state charges are taken 
into account. Parameter choices are explained in the text.}

\bigskip

\normalsize
It is useful to note that ($a$) for a given selectron and a given coupling,
the sum of the cross-sections in the three columns corresponds to the
cross-sections shown in Figs. 1 and 2 (diminished by the
application of cuts), hence it is trivial to calculate the branching
ratios to the different channels; and ($b$) the above
cross-sections have been calculated without any cuts on the missing $E_T$.
However, a cut of $\mET > 20$ GeV does not lead to any significant change
(see below) in the cross-sections for this value of LSP mass (93.5 GeV),
so these numbers may be considered representative.
The principal SM background to these various signals will come from
processes like $\egam \goes e Z W^+ W^-,
\nu Z Z W^-$, followed by the leptonic decays of the $W,Z$ bosons.
It is straightforward to estimate \footnote{For example,
the backgrounds presented in Ref. \cite{Barger} would be
further suppressed by the fine structure constant and then by the leptonic
branching ratios.} that such cross-sections
are very small indeed (typically $\sim 1$ fb $\times$ relevant
leptonic branching ratios) and should not constitute a source of
serious worry. We do not consider these backgrounds further in the
current study.

\begin{figure}[h]
\begin{center}
\vspace {5.0in}
      \relax\noindent\hskip -4.0in\relax{\includegraphics{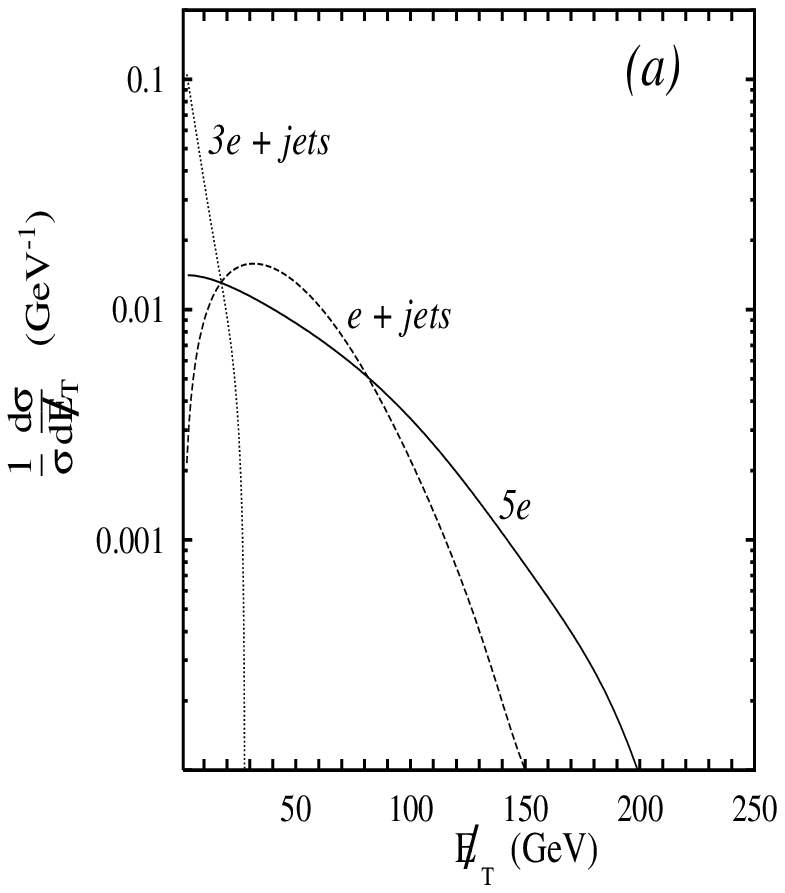}}
      \relax\noindent\hskip  2.3in\relax{\includegraphics{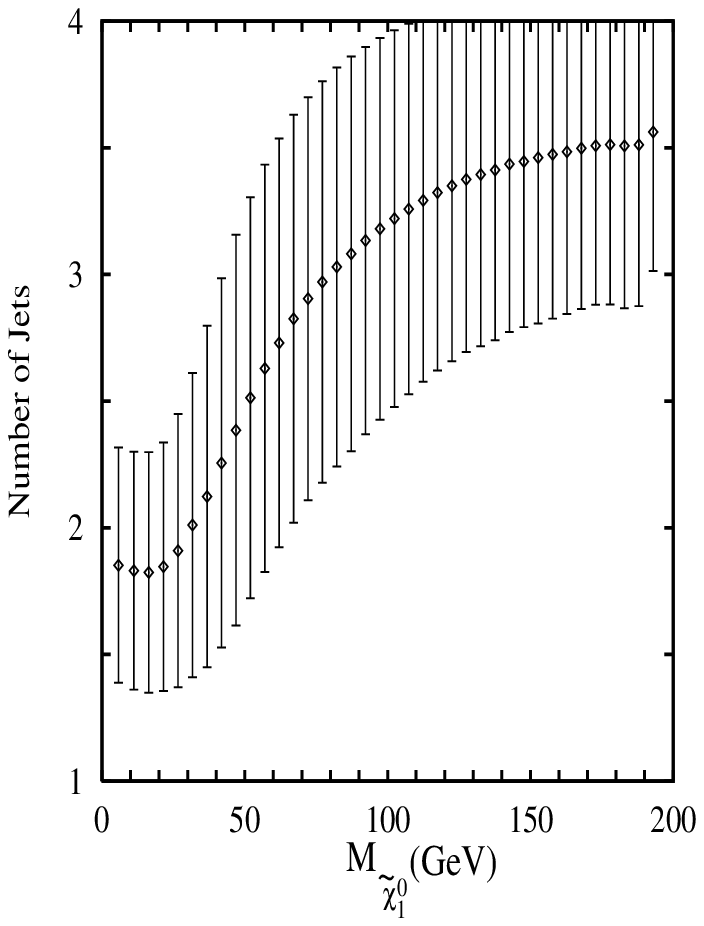}}
\end{center}
\vspace*{-120pt}
\caption{\footnotesize\em 
Kinematic profiles for the $R$-parity-violating signals. (a) Missing 
transverse energy distributions, and (b) variation of the average number
of jets for a $\lambda'_{121}$ coupling and a final state with  
$e + \mET$ + jets.}
\end{figure}

As mentioned above, the charged leptons in the final state will
always be accompanied by substantial missing energy and momentum.
In Fig. 4($a$), the solid line shows the missing (transverse) energy
distribution for a $\l_{121}$ coupling and a 5-electron final
state. Parameter choices coincide with Table 1.
A substantial missing energy, say $\mET > 20$ GeV, should be
considered an important part of the signal, and
Fig. 4($a$) shows that this would affect the signal only marginally.
Of course, the missing energy is rather sensitive to the neutralino
mass, so that very light neutralino masses might then
be excluded from this
analysis. These, however, would be accessible to searches in $\epem$
collisions at LEP and at the NLC itself.
 
We now turn to the case of $LQ\bar D$ operators and $\lp$ couplings, 
of which three ($\lp_{121},\lp_{131},\lp_{123}$) are of interest in 
possible explanations of the HERA excess \cite{HERA_RpV}. 
In the presence of a $\lp_{ijk}$ coupling, the neutralino
undergoes the three-body decays 
\begin{displaymath}
\N0_1 \goes \nu_i d_j \bar d_k + \ell_i^- u_j \bar d_k
\end{displaymath}
provided they are kinematically allowed,
through a virtual $\snu_{i,j}, \sl_{i,j}, \sd_{j,k}, \su_j$
exchange. This leads to the set of final states listed in Table 2.
Only the leptonic content of the final state is shown; this will always
be accompanied by hadronic activity. As before, the last column contains
a combinatoric factor of 2.

\bigskip

\footnotesize
\begin{center}
\begin{tabular}{|c|c|c|c|c|c|c|}
\hline
$\lp$ & Final state & C.S.(fb) & Final state & C.S.(fb) 
&  Final state & C.S.(fb)    \\
\hline
$11k, 12k$ & $ ~e^- + \mET$         & 7.03 (17.50) &
             $ ~e^- + e^\pm e^\pm$  & 2.27 (5.60) &
             $ ~e^- + e^\pm + \mET$ & 8.22 (20.47)    \\
\hline
$i3k$      & $ ~e^- + \mET$          & 44.45 (110.78) & & & & \\
\hline
$21k, 22k$ & $ ~e^- + \mET$           & 7.02 (17.50) &
             $ ~e^- + \mu^\pm \mu^\pm$ & 2.28 (5.60) &
             $ ~e^- + \mu^\pm + \mET$ & 8.24 (20.46)    \\
\hline
$31k, 32k$ & $ ~e^- + \mET$          & 7.05 (17.58) & 
             $ ~e^- + \tau^\pm \tau^\pm$  & 2.71 (6.47) &
             $ ~e^- + \tau^\pm + \mET$ & 8.24 (22.79)   \\
\hline
\end{tabular}
\end{center}
\noindent
{\em Table 2. Final states for different $\lp$ couplings with
representative cross-sections for each channel. Notations and 
parameter choices are the same as in Table 1. }
\normalsize

\bigskip

Each final state will, in fact, contain $n = 1,2,3,4$ hadronic jets 
(depending on whether the jets merge or not). The cuts imposed in order to
get the above cross-sections are (as before) 
$|\eta_\ell| < 3$ and $E_T^\ell > 10$ GeV
for each charged lepton. These hardly affect the signal 
for a LSP mass as heavy as 93.5 GeV. For lower masses, of course, the
$\mET$~cut would become significant. Since the final state
is rather messy, with up to four hadronic jets, one has also to impose
isolation criteria on all the leptons including narrow $\tau$-jets, for 
which we require $\Delta R_{\ell j} > 0.4$ (0.55 for $\tau$ jets). 
These further reduce the cross-section by about 37\% (47\% if there are
$\tau$'s.)  As before, we assume efficiencies
of 95 (80)\% in the identification of each $e,\mu(\tau)$. Despite these
suppression factors, it  seems clear that we still predict significant
numbers of events for 10 fb$^{-1}$ luminosity.

For the jets, we demand $|\eta_j| < 3$ and $E_T^j > 15$ GeV and assume 
that jets with an angular separation $\Delta R_{jj} < 0.7$ merge into
a single jet. The rapidity and minimum energy criteria add to the missing
energy of the event, but this does not affect the numbers in Table 2,
since no cut on missing energy was used in generating them. The missing 
(transverse) energy
distribution for a final state with 3 electrons + jets is plotted as a dotted
line in Fig. 4($a$). Since there are no neutrinos in this final state
the missing energy arises solely from jets and hence this may be taken
as an indicator of the minimum missing energy required to identify
neutrino final states. 
 
One of the more spectacular possibilities in the case of $\lp$ couplings
is the case of two or even three like-sign leptons in the final state.
Though one has to pay a price of a factor of one-half (one-quarter) in
cross-section to observe like-sign dilepton (trilepton) signals, the 
cross-sections are just about large enough and there
is no SM background worth consideration. Such a signal is the 
surest sign of a Majorana particle in the decay chain and
is likely to play a major role in the identification of neutralinos
in a model with $R$-parity violation.

In the above we assume that the neutralino LSP cannot decay into a top
quark, which accounts for the columns left blank in Table 2 and the
fact that the branching ratio for neutralinos to neutrinos and jets is
unity for $\lp_{i3k}$. 
This is certainly true for the parameter choice of Tables 1 and 2
for which the neutralino mass is 93.5 GeV.
Should the LSP be heavier than the top quark, the
final states for a $\lp_{i3k}$ coupling would involve the top quark
and its decay products. For the energies
assumed at the NLC, however, the production of neutralinos much
heavier than the top quark is suppressed, so that 
neutralinos produced in $\egam$ collisions with any significant 
cross-section will have masses rather close to the top quark mass,
in which case, the neutralino branching ratio to the top quark will
be kinematically suppressed and this decay channel will
not be competitive. 
The corresponding Cabibbo-Kobayashi-Maskawa-suppressed
decay channel involving a charm quark is also negligible compared with
the neutrino plus jets channel. The whole situation changes 
somewhat if a higher energy machine is contemplated, but that will 
not concern us in the present work.

Apart from the dramatic multilepton signals, the case of an electron
accompanied by jets and missing energy is also worth serious
consideration, since such signals are
predicted irrespective of which $\lp$ coupling is involved. In fact,
two of the couplings of interest for the HERA excess, \viz, $\lp_{131}$
and $\lp_{132}$ can be accessed only through this mode. In this case
signal cross-sections can be large for $\lp_{i3k}$ couplings, but 
we expect larger SM backgrounds too from processes like $\egam
\goes eZZ,~\nu ZW$, followed by hadronic decay of
one of the $W,Z$ bosons (for which the branching ratios are higher)
with gluon radiation making up the tally of jets.
However, these backgrounds should be manageable
with judicious cuts. The dashed line in 
Fig. 4($a$) shows the missing (transverse) energy distribution
for the signal 
assuming a $\lp_{121}$ coupling and a final state with $e^-$, jets and
missing momentum. Parameter choices coincide with Table 2. 
As in the case of purely leptonic
final states, a cut of $\mET > 20$ GeV does not hurt the signal, though 
it must again be admitted that 
the missing energy will be rather soft if the
neutralino is light. 

In Fig. 4($b$), we show a plot of the number of jets in the 
$e$ + $\mET$ + jets final state against neutralino mass for a 
left selectron with mass 200 GeV. 
The central value shows the mean number of jets and the bars show the 
statistical spread obtained in a simulation with 50000 events. 
The number of jets is determined almost wholly by the kinematics 
and has very little dependence on the other parameters of the model.
It is apparent that for light neutralinos below 50 GeV, one would tend
to have two jets, which is simply because the neutralinos are highly 
boosted and their decay products tend to lie in a narrow cone about
the original direction. In this case, we will undoubtedly have a SM
background from $\egam \goes e q \bar q$ through a $Z$ resonance to
worry about, and one would probably 
require a strong cut on the missing energy and maybe a cut 
on the invariant mass of the jets to remove the effect of the $Z$-resonance. 
The missing energy criterion alone should ensure that
the background becomes negligible. 
Moreover, as the mass of the neutralino, increases,
the final state would tend to have three to four jets.
For the SM background, this requires radiation of at least one gluon and 
hence suppression of the cross-section by one or more powers of the strong 
coupling constant $\alpha_s$.

Finally, we briefly discuss the case when baryon-number is violated.
In this case, each neutralino (LSP) would
decay into three jets, so that the final signal would be a
hard, isolated electron and up to six hadronic jets. In the absence of 
multileptons in the final state it is essential to trigger on 
the sole electron which must then 
satisfy energy and angular criteria as defined above, and, in addition,
must be isolated from {\em all} the hadronic jets. Given the wide spatial 
distribution
of hadronic debris, this criterion would plainly lead to a significant 
reduction in the cross-section. Backgrounds from
$\egam \goes e Z \goes e q \bar q$, followed by gluon 
radiation, can perhaps be controlled as before, though there is no
simple criterion such as missing energy to distinguish signal from 
background. 
A detailed simulation of the multijet production and merging would be 
needed, therefore,
and suitable cuts devised, in order to isolate the signal.
Such an analysis is beyond the scope of the present work.

To conclude, then, we have explored the possibility that an $\egam$
collider can produce a selectron in association with a neutralino
(LSP). Decay of the selectron can yield a final state with an
electron and two neutralinos. We assume that $R$-parity is weakly 
violated
and thus the neutralinos will decay into three-fermion states.
Different possibilities have been considered and it seems that rather
optimistic discovery limits can be set for lepton-number-violating
couplings. Should an $\egam$ collider be built, therefore, one can 
look forward to significant advances in the study of supersymmetry, 
including the option of $R$-parity violation considered here.

The authors are grateful to S.~Bhattacharyya (L3 Collaboration), 
D.~Choudhury and R.M.~Godbole
for discussions. SR acknowledges partial financial support from 
the World Laboratory, Lausanne, and the hospitality of the Theory Group
of Fermilab, where a part of this work was done. The work of DKG is 
funded by the University Grants Commission, Government of India and 
he thanks the Centre for Theoretical Studies, Indian Institute of
Science, Bangalore, for hospitality.

\end{document}